\documentclass[prl,twocolumn]{revtex4}
\usepackage{bm}
\usepackage{graphicx}
\usepackage{amsmath}
\usepackage{eufrak}
\usepackage{color}
\usepackage{upgreek}
\usepackage{subfigure}
\usepackage[unicode=true,colorlinks=true,citecolor=blue]{hyperref}
\usepackage{placeins}
\usepackage{lipsum}

\newcommand{\nix}[1]{}

\begin{document}

\title{High-frequency rectification in graphene lateral $p$-$n$ junctions}

\author{Yu. B. Vasilyev,$^1$ G. Yu. Vasileva,$^1$ S. Novikov,$^2$
S. A. Tarasenko,$^1$  S. N. Danilov,$^3$ and S. D. Ganichev,$^3$
}
\affiliation{$^1$Ioffe Institute, 194021 St.\,Petersburg, Russia}

\affiliation{$^2$ Micro and Nanoscience Laboratory, Aalto University, Tietotie 3, FI$n$-02150, Espoo, Finland}

\affiliation{$^3$Terahertz Center, University of Regensburg, 93040 Regensburg, Germany}

\begin{abstract}
We observe a \textit{dc} electric current in response to terahertz
radiation in lateral inter-digitated double-comb graphene $p$-$n$
junctions. The junctions were fabricated by selective ultraviolet
irradiation  inducing $p$-type doping in intrinsic $n$-type
epitaxial monolayer graphene. The photocurrent exhibits a strong
polarization dependence and is explained by electric rectification
in $p$-$n$ junctions.
\end{abstract}

\maketitle


The development of graphene electronic circuits is one of the
important tasks of modern solid-state electronics. Doping or
electrical gating of graphene enable the fabrication of $p$-$n$
junctions, which are the subject of intensive study, see
e.g.~\cite{1,2,5,3,4}. It was found experimentally that graphene
structures with lateral $p$-$n$ junctions demonstrate almost
symmetric current-voltage ($I$-$V$) characteristics, which are
very similar for the negative and positive polarities of the
source-drain voltage~\cite{1,2,5,3}. The bi-directional charge
transport is attributed to the Klein tunneling of Dirac
fermions~\cite{6}. The symmetric character of the $I$-$V$
characteristic results in a lack of a pronounced rectification of
electric signals in graphene $p$-$n$ devices being an obstacle for
their application in electronics.

In optical measurements, however, it was observed that the
illumination of graphene $p$-$n$ junctions leads to an electric
response, which is a prerequisite for the fabrication of
photodiodes~\cite{8,9}. At high photon energy
the photoresponse is caused by the creation of electron-hole
pairs, which are separated by a built-in electric field of the
$p$-$n$ junction~\cite{10}. This mechanism, however, fails for
lower frequencies of terahertz (THz) range, for which  the photon
energy is typically much lower than the Fermi energy and the
generation of electron-hole pairs is blocked. Nevertheless,
irradiating sharp lateral graphene $p$-$n$ junctions by THz laser
radiation we observe a $dc$ current. The photocurrent is detected
in epitaxial graphene on a SiC substrate with inter-digitated
dual-comb $p$-$n$ junctions and is characterized by a strong
dependence on the radiation polarization. The observations suggest
the rectification of THz electric field in graphene  $p$-$n$
junctions.

\begin{figure}[t]
\includegraphics[width=\linewidth]{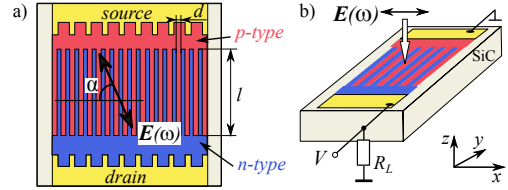}
\caption{ (a) Top view of
graphene photodiode, where the source and drain electrodes
(yellow) are in contact with the $p$- (red) and $n$-type (blue)
areas.
(b) Experimental setup. }
\label{fig_1}
\end{figure}

Epitaxial graphene samples for the subsequent $p$-$n$ junction
fabrication were prepared by high-temperature Si sublimation of
SiC~\cite{11}. In order to reduce the density of carriers we
exposed the prepared samples in hot air~\cite{25}. This resulted
in the initial $n$-type carrier density in the range $(1.6 \div
3.1)\times 10^{11}$ cm$^{-2}$ and the mobility of about $(1400
\div 2700)$~cm$^2$/(V s). Then, the graphene was covered with
polymers
PMMA
and
ZEP500.
This enable the adjusting of the carrier density by photochemical
gating in which ZEP500 provide potent acceptors under deep
ultraviolet (UV) light~\cite{26}. By irradiation with UV through
the shadow mask graphene was patterned into $p$-$n$ structures
consisting of inter-digitated finger stripes of $p$- and
$n$-types, see Fig.~\ref{fig_1}. The irradiation dose was chosen
high enough to inverse the carrier type from $n$- to
$p$-type~\cite{12}. Consequently, $p$-$n$ junctions are formed
along the boundary of the illuminated and not illuminated areas.
Inter-digitated devices with finger length $l=1.8$~mm and width of
$d=50$ and 100~$\mu$m for samples $\# A$ (68 junctions), and $\#
B$ (34 junctions), respectively, were fabricated to enlarge the
area of the $p$-$n$ junctions. The type and degree of doping in
the fingers were extracted from the Hall measurements. The
obtained carrier densities are $3\times 10^{11}$~cm$^{−2}$ for
the $n$-doped part and $5\times 10^{10}$~cm$^{−2}$ for the
$p$-doped part. The corresponding Fermi energies $E_F$ are 70 and
25~meV, respectively.


The experiments on photocurrents are performed applying radiation
of NH$_3$~laser~\cite{Ganichev93,Schneider04} operating at the
frequencies $f=2$ and 1.1~THz. We use single pulses with a peak
power of $P \approx 10$~kW, a duration of about 100~ns  and a
repetition rate of 1~Hz. The beam had an almost Gaussian shape, as
measured by a pyroelectric camera~\cite{LechnerAPL2009},
and was focused at a spot with the diameter $\approx 2$~mm. All
experiments are performed at normal incidence of radiation,
Fig.~\ref{fig_1}(b), and
temperature $T = 4.2$~K.
We use linearly polarized radiation with the direction of the
radiation electric field $\bm E$ relative to the $p$-$n$ junction
described by the  angle $\alpha$, see Fig.~\ref{fig_1}(a).  The
latter was varied applying $\lambda/2$-plates~\cite{15a}.

\begin{figure}[t]
\includegraphics[width=\linewidth]{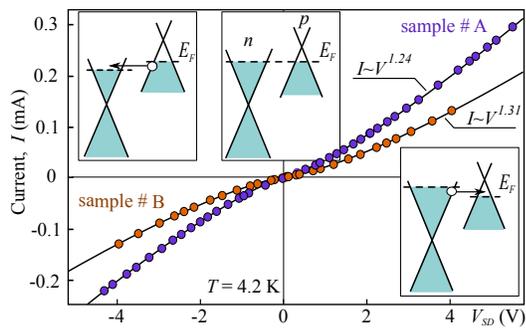}
\caption{ $I$-$V$ characteristics.
Insets illustrate the
mechanism of the current flow in direct- and reverse-biased $p$-$n$ junctions.}
\label{fig_2}
\end{figure}

First, we study $I$-$V$ characteristics
applying $dc$ voltage, $V_{SD}$, between the source and drain contacts,
see Fig.~\ref{fig_2}. The data reveal that there is no asymmetry
in the behavior of the current for the forward- and reverse-biased
junctions. The symmetry indicates that our graphene structures as
a whole do not exhibit rectifying properties, which is in contrast
to conventional diodes based on gaped semiconductors. A
substantial current flow at the reverse-biased $p$-$n$ junction is
attributed to the absence of a band gap in graphene and the
related efficient Klein tunneling~\cite{6}. The corresponding band
diagrams, sketched in Fig.~\ref{fig_2}, show that the electric
current can flow for both polarities of the applied voltage.

The ability of a graphene $p$-$n$ junction to conduct  electric
current in both directions, however, implies neither that the
$I$-$V$ characteristic  itself is symmetric nor that the $p$-$n$
junction cannot rectify $ac$ electric fields. In fact, the real
$I$-$V$ characteristic of the $p$-$n$ junction and its possible
asymmetry can be hidden in transport experiments and its
measurement in the lateral graphene $p$-$n$ junction is a
challenging task. This is because the current in such samples is
determined by the large resistance of the $p$- and $n$-type areas
rather than the small resistance of the thin $p$-$n$ junction.
Note, in vertical graphene $p$-$n$ junctions, where the lateral
resistance of the $p$-type and $n$-type areas does not play a
role, an asymmetry of the $I$-$V$ characteristic was indeed
observed~\cite{16,17}.

The conclusion that the $I$-$V$ characteristics in our  lateral
graphene samples are determined by the $p$- and $n$-type areas is
supported by the fact that they
are well described by the dependence $I\propto V^\delta$  with
$1<\delta <1.5$, see  Fig.~\ref{fig_2}. Similar superlinear
dependence was obtained earlier in experiments on homogeneous
graphene samples and explained by the interplay of the intraband
and interband contributions to the charge transport~\cite{18}. The
dependence $I\propto V^{1.5}$ also follows from theoretical
calculations in the ballistic regime, see Refs.~\cite{19,20}.

\begin{figure}[t]
    \includegraphics[width=\linewidth]{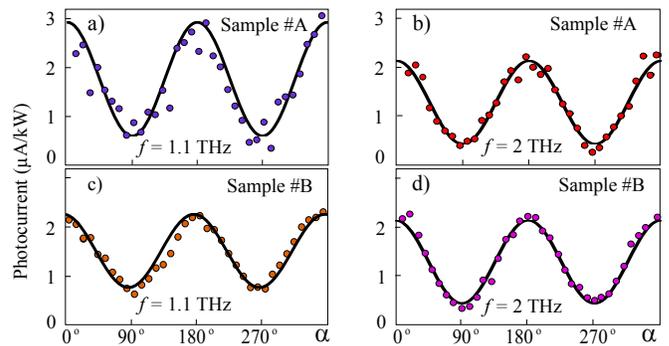}
\caption{
 Dependence of the photoresponse on the angle $\alpha$ which
 determines the in-plane orientation of the THz electric field.
 The photoresponse is measured at the normal incidence of
 radiation in two samples for two different wavelengths.
}
    \label{fig_3}
\end{figure}

To measure  the response of the lateral $p$-$n$ junctions to a
high-frequency electric field we excite the structure by THz
radiation in the absence of an applied bias voltage (photovoltaic
mode). Illuminating the structure at normal incidence we detect a
\textit{dc} photocurrent between the source and drain contacts
which linearly scales with the radiation intensity and has a
strong polarization dependence, see Fig.~\ref{fig_3}. The fact
that the photoresponse emerges at normally incident radiation
unambiguously indicates that it originates from the $p$-$n$
junction. Indeed, all known mechanisms of the photocurrent
formation in homogeneous graphene structures, such as photon drag
or photogalvanic effect, require the oblique incidence of
radiation or breaking the symmetry by a magnetic
field~\cite{jiangPRB2010,22}. The photocurrent also cannot be
caused by ratchet~\cite{Olbrich16} or plasmonic~\cite{10} effects
recently observed in graphene  because they require
metal gates superlattices.

The role of the $p$-$n$ junction  also follows from the dependence
of the photoresponse on the angle $\alpha$ between the electric
field of the radiation and the normal to the $p$-$n$ junctions,
see ~\ref{fig_3}. For both wavelengths studied the photocurrent
reaches a maximum at $\alpha = 0$ and $180^\circ$, when the $ac$
electric field of the incident radiation is perpendicular to the
$p$-$n$ junctions, and a minimum at  $\alpha = 90^\circ$, when the
radiation is polarized along the $p$-$n$ junctions. The overall
polarization dependence
is well fitted by
 \begin{equation} \label{FITalpha}
J(\alpha) =  (a + b \cos^2{\alpha}) P \:.
 \end{equation}
Note that the observed variation of the photoresponse
is very large so that
$(J_{\rm max} - J_{\rm min})/(J_{\rm max} + J_{\rm min}) = 0.8 \div 0.9$
being close to unity.
Furthermore, increasing the number of $p$-$n$ junctions by
decreasing the width of the fingers in the combs (samples $\# A$
vs. $\# B$) we observed an increase of the signal for low
radiation frequencies. Comparing the photocurrent excited by
$f=1.1$ and 2~THz radiation we see that in sample $\# A$ it
decreases with raising frequency whereas in sample $\# B$ it
remains unchanged.

Now we discuss the origin of the photoresponse. Since the energy of photons of THz radiation
is much smaller than the Fermi energies of carriers in  $p$-type and $n$-type areas, as well as in $p$-$n$ junction areas, the interband absorption of radiation with the creation of electron-hole pairs is blocked.
%
In this case, the Drude absorption dominates and the
quasi-classical approach to the description of electron transport
through the $p$-$n$ junction becomes valid. The observed $dc$
photocurrent can be interpreted as a result of the rectification:
The $ac$ electric field of the THz radiation incident upon  the
$p$-$n$ junction causes a high-frequency electric current, which
is partially rectified due to the asymmetry of the $I$-$V$
characteristic of the $p$-$n$ junction.
The rectification occurs in the area of the $p$-$n$ junction while,
as discussed above, the homogenous $p$- and $n$-doped parts 
do not contribute to the formation of the $dc$ signal. To the
lowest order in the electric field amplitude, the $dc$
current
is described by the second order nonlinear term in the $I$-$V$ characteristic:
%
\begin{equation}
\label{FITalpha}
I_{dc} =  \sigma_2 E^2_{\perp} \:,
\end{equation}
where $\sigma_2$ is the nonlinear conductance and  $E_{\perp} $ is
the amplitude of the $ac$ electric field across the $p$-$n$
junctions. The
photocurrent
scales quadratically with the electric field amplitudes as
detected in experiment.
The rectification mechanism also explains
the observed polarization dependence.
The $dc$ electric current is generated
by the component of the $ac$ electric field normal to the $p$-$n$ junction
varying as
$E^2_{\perp} \propto \cos^2(\alpha) $   which is in agreement with
the experimental data, Fig.~\ref{fig_3}. The presence of an
additional small polarization-independent signal may be related to
the photo-thermoelectric effect~\cite{24}.

The observed weak frequency dependence  of the photosignal in
sample $\# A$ and its absence in the sample $\# B$ are due to low
mobility of carriers in our samples. Indeed, the frequency
dependence is determined by the parameter $\omega \tau$, where
$\omega = 2 \pi f$ and $\tau$ is the momentum relaxation time. For
the carrier density $3\times 10^{11}$~cm$^{-2}$, mobility
$10^3$~cm$^2$/(Vs), and the radiation frequency $f=1$~THz, the
estimation yields $\omega \tau \approx 0.05$. It shows that the
$ac$ transport of carries induced by THz radiation in the $p$-$n$
junctions is, in fact, quasi-stationary and depends on the $ac$
electric field amplitude rather than on its frequency.

In conclusion, we have reported photocurrent  measurements in
graphene lateral $p$-$n$ junctions formed by selective deep UV
illumination the graphene monolayer fabricated on a SiC substrate.
We have observed a pronounced photoresponse in the THz range,
which has a strong dependence on the radiation polarization. The
observations are explained by the rectification mechanism of the
photocurrent formation in graphene $p$-$n$ junctions.


We thank V.
Belkov
and V.
Kachorovskii for fruitful
discussions. The support from the RFBR
(16-02-00326), and the DFG
(SFB~1277-A04)
is acknowledged.
S.T. acknowledges the support from the RSF (14-12-01067).%
%
%


\end{document}